# Autoencoder based Domain Adaptation for Speaker Recognition under Insufficient Channel Information

*Suwon Shon*[1], *Seongkyu Mun*[2], *Wooil Kim*[3], *Hanseok Ko*[1],

[1]School of Electrical Engineering, Korea University, South Korea
[2] Dept. of Visual Information Processing, Korea University, South Korea
[3] Dept. of Computer Science and Engineering, Incheon National University, South Korea

swshon@korea.ac.kr, hsko@korea.ac.kr

## Abstract

In real-life conditions, mismatch between development and test domain degrades speaker recognition performance. To solve the issue, many researchers explored domain adaptation approaches using matched in-domain dataset. However, adaptation would be not effective if the dataset is insufficient to estimate channel variability of the domain. In this paper, we explore the problem of performance degradation under such a situation of insufficient channel information. In order to exploit limited in-domain dataset effectively, we propose an unsupervised domain adaptation approach using Autoencoder based Domain Adaptation (AEDA). The proposed approach combines an autoencoder with a denoising autoencoder to adapt resource-rich development dataset to test domain. The proposed technique is evaluated on the Domain Adaptation Challenge 13 experimental protocols that is widely used in speaker recognition for domain mismatched condition. The results show significant improvements over baselines and results from other prior studies.

## 1. Introduction

The i-vector extraction paradigm, which utilizes the Universal Background Model (UBM) and total variability factor space, has sparked great interest in the field of speaker recognition due to its remarkable performance over the past few years [1]. The i-vector length normalization and Probabilistic Linear Discriminant Analysis (PLDA) approaches have been employed to effectively score between test utterances and speaker models [2]–[6].

For best performance, the domain of the data used in system development should match the domain of the system that will be operated. In this case, if a dataset domain is matched with operation domain, it is called *in-domain* dataset; if it is mismatched, it is called *out-of-domain* dataset. Garcia-Romero [7] found that training the UBM and total variability subspace on some types of domains have limited effects on performance improvement. Instead, the performance depends heavily on PLDA parameter estimation. By estimating PLDA parameters on large labeled in-domain dataset, the system could improve its performance and it performed almost as a system trained on the in-domain itself.

However, large labeled in-domain datasets are not generally available or they are highly expensive to purchase. Thus, in practice, speaker recognition systems usually suffer performance degradation when applied to unknown domains. To be generally applicable, speaker recognition systems must overcome this performance limitation which comes from the insufficient information of in-domain. Solutions to domain mismatched condition proposed by others can be divided into two general scenarios depending on the particular information availability scenarios they are designed for. Both scenarios assume that a large out-of-domain dataset is available with speaker labels.

Under the first scenario, corpus label is available for the out-of-domain dataset and in-domain dataset is not available. Although the system lacks in-domain information, the system can compensate for the domain mismatch by a corpus matched whitening transformation with a whitening library [8]. Furthermore, it also uses corpus dependent subspaces employing the techniques of Within-class Covariance Correction (WCC) [9] and Inter-Dataset Variability Compensation (IDVC) [10], [11],.

Under the second scenario, an unlabeled in-domain dataset is available. However, it lacks speaker labels, so it cannot be used directly with the PLDA parameter estimation. Villalba [12] introduced a variational Bayesian approach for adapting PLDA models from out-of-domain datasets to in-domain. Garcia-Romero [13] introduced a clustering approach for unlabeled in-domain datasets that uses well-known Agglomerative Hierarchical Clustering (AHC) to estimate Within-speaker Covariance (WC) and Across-speaker Covariance (AC) with the PLDA model. In this approach, the estimated in-domain WC and AC are interpolated from out-of-domain WC and AC [7], [13], [14]. Kanagasundaram [15] introduced another IDVC technique, called Inter-Dataset Variability (IDV), to capture the variability between out-of-domain and in-domain. Furthermore, Rahman [16] proposed a Dataset-Invariant Covariance Normalization (DICN) approach which is similar to IDV, but this approach is more effective in compensation than IDV. Denoising Autoencoder (DAE) based approach also can be used. Autoencoder based domain adaptation is widely used in machine learning community [17]–[19] and also adopted already on speech processing area [20]–[22]. Recently Kudashev [23] proposed a DAE-based denoising and domain adaptation for speaker recognition.

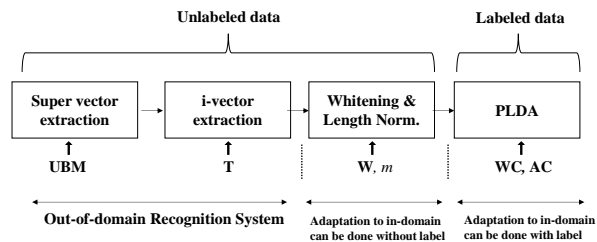

Figure 1: *Block diagram of speaker recognition system*

Prior studies on the second scenario assume that the in-domain dataset is sufficiently resource-rich to estimate within-speaker and across-speaker variability. However, if the in-domain dataset lacks sufficient audio samples of each speaker to estimate within-speaker variability, the dataset cannot be effectively used for adaptation. This study explores the performance degradation in such a situation, *insufficient channel information*.

Because the small set of in-domain dataset cannot be used directly to estimate PLDA parameters by lack of speaker information, there is a need for new approach. In this paper, the use of Autoencoder based Domain Adaptation (AEDA) is proposed in order to cope with insufficient channel information of in-domain dataset on second scenario. AEDA combines an Autoencoder (AE) with a DAE to adapt the resource-rich out-of-domain dataset utilizing in-domain dataset. Training the AE part uses both in-domain and out-of-domain i-vectors and training the DAE part uses sparse reconstructed out-of-domain i-vectors using in-domain dataset dictionary. In this paper, the performance of this new approach is compared to conventional domain adaptation and compensation approaches.

The rest of the paper is organized as follows. Sec. 2 outlines the baseline speaker recognition system and describes domain mismatch conditions under insufficient channel information. Sec. 3 presents the proposed unsupervised domain adaptation approach, and Sec. 4 presents and analyzes the experimental results. Sec. 5 concludes the paper.

## 2. Speaker Recognition in Mismatched Conditions

### 2.1. Speaker Recognition System and Adaptation

Fig.1 is a block diagram of the i-vector based speaker recognition system with the parameters needed for each process. The first block indicates the UBM and the second indicates the training total variability matrix **T** which takes advantage of a large out-of-domain dataset. The third block is the pre-processing step that satisfies the i-vector as a Gaussian modeling assumption of fourth block through whitening and length normalizing. The fourth block scores input utterances from the speaker model using the PLDA parameters. When adapting the system to in-domain, parameters of the third and fourth block must be estimated on the in-domain.

### 2.2. Adaptation under Insufficient Channel Information

We conducted experiments using Speaker Recognition Evaluation (SRE) and Switchboard (SWB) dataset which are widely used by in speaker recognition community. On this paper, SWB and SRE are considered as group of i-vector sets. The SWB set consists of all telephone calls from all speakers taken from SWB 1 and 2(phase 1 to 3), and SWB 2 Cellular phase 1 and 2. The SRE set is taken from the SRE 04, 05, 06 and 08 collections. For domain mismatched condition, Domain Adaptation Challenge 13 (DAC13) experimental protocol is used [24]. DAC13 provides the list and i-vectors from SRE and SWB dataset that were used as table I.

Table 1: *i-vector Statistics in DAC 13 i-vector Dataset*

|  | SWB | SRE | SRE-1phn |
|---|---|---|---|
| #spkrs | 3114 | 3790 | 3787 |
| #calls | 33039 | 36470 | 25640 |
| Avg. #calls/spkrs | 10.6 | 9.6 | 6.77 |
| Avg. #phone_num/spkr | 3.8 | 2.8 | 1 |

Table 2: *SRE10 Test using DAC13 i-vector set.*

| System # | **UBM**, **T** | **W**, *m* | **WC**, **AC** | EER |
|---|---|---|---|---|
| 1 | SWB | SRE | SRE | 2.33 |
| 2 | SWB | SRE | SWB | 5.70 |
| 3 | SWB | SRE-1phn | SRE-1phn | 9.34 |
| 4 | SWB | SRE-1phn | SWB | 5.66 |

Additionally, DAC13 provided the SRE-1phn i-vector set, which is reduced set consisting of the i-vector from only 1 telephone number per speaker. Since DAC13 established a benchmark for domain adaptation evaluation by setting SWB as the development out-of-domain dataset and SRE as the evaluation in-domain dataset, many studies explored domain mismatch issued. However, comparisons of performance under insufficient channel information have been inadequate.

Table II shows the Equal Error Rate (EER) performance of the SRE10 c2-extended test [24] when the parameters are trained with different datasets using the framework presented in Fig. 1. System 1 can be considered as the desired benchmark when the in-domain dataset speaker label is known. System 2 is the baseline of the domain mismatched condition when the in-domain database is unlabeled. Systems 3 and 4 are versions of systems 1 and 2, respectively on more challenging conditions. System 3 is adapted using a matched in-domain labeled dataset SRE-1phn which is subset of SRE. Note that, although system 3 is under domain matched conditions and system 4 is under mismatched conditions, system 4 shows better performance in EER than system 3. This is an interesting result and we believe that performance was degraded by *insufficient channel information*. The dataset 'SRE-1phn' contains audio from only a single telephone number per speaker and use of such a poor phone number diversity hinders the effective estimation of within-speaker variability of in-domain. In this case, the conventional approaches [13], [14] that estimate within-speaker variability from in-domain unlabeled dataset would fail, in spite of perfect speaker label estimation, due to insufficient channel information. Singer [25] also tackled same issue and suggested dataset selection criteria to prevent this situation in advance.

## 3. Proposed Unsupervised Domain Adaptation

### 3.1. Autoencoder and Denoising Autoencoder

AE is an artificial neural network for reconstructing their own input signal. DAE is the recent variant of the classical AE for reconstructing clean repaired inputs from corrupted input signals. Both encoders have an output layer featuring the same number of nodes as the input layer and one or more hidden layers connecting those input and output layers. The nonlinearity of the encoder and decoder ensures feasibility of the denoising approach and shows promising results when searching for more robust features for speech enhancement [21]. This study uses AE and DAE, which contain a single hidden layer.

### 3.2. Proposed Autoencoder-based Domain Adaptation

In this paper, the idea of DAE is extended to AEDA by replacing denoising concept with domain adaptation. Suppose that there is autoencoder that adapts out-of-domain dataset to in-domain using only unlabeled in-domain dataset. Then, resource-rich out-of-domain dataset could be more useful for speaker recognition system than using unlabeled in-domain dataset that degrades performance due to insufficient channel information as discussed in Sec.2.B. This scheme is shown in Fig. 2 where in-domain i-vector set is $D_{in}$ and out-of-domain i-vector set is $D_{out}$. $D_{out}^{t}$ is adapted i-vector set from $D_{out}$.

Fig. 3 depicts the proposed AEDA structure which combines AE and DAE with sharing decoder $g$. The AE part consists of the encoder $f_{in}$ and decoder $g$. the DAE part consists of the encoder $f_{out}$ and decoder $g$. The $f_{in}$ and $f_{out}$ encoders map both domain inputs to the hidden representation $\mathbf{h}$ as $f_{in}(\mathbf{x}_{in}) = \sigma(\mathbf{W}_{in}\mathbf{x}_{in} + \mathbf{b}_{in})$ and $f_{out}(\mathbf{x}_{out}) = \sigma(\mathbf{W}_{out}\mathbf{x}_{out} + \mathbf{b}_{out})$ where $\sigma$ is an element-wise non-linear activation function, such as a sigmoid function. $\mathbf{W}_{in}$ and $\mathbf{W}_{out}$ are weight matrices $\in \mathbf{R}^{n \times m}$ where $n$ is the number of hidden units and $m$ is the number of input layer nodes. $\mathbf{b}_{in}$ and $\mathbf{b}_{out}$ are bias vectors $\in \mathbf{R}^{n}$. The sharing decoder $g$ maps the hidden representation $\mathbf{h}$ to in-domain output $\mathbf{Y}_{in}$ as $g(\mathbf{h}) = \mathbf{W}\mathbf{h} + \mathbf{b}$ where $\mathbf{W} \in \mathbf{R}^{m \times n}$ and $\mathbf{b} \in \mathbf{R}^{m}$ are the decoder parameters. The objective function of the AE part is

$$\min_{f_{in}, g} \| \mathbf{X}_{in} - \mathbf{Y}_{in} \|^{2} = \min_{f_{in}, g} \| \mathbf{X}_{in} - g(f_{in}(\mathbf{X}_{in})) \|^{2} \quad (1)$$

where $\mathbf{X}_{in}$ is the in-domain inputs from $D_{in}$. The objective function of the DAE part is

$$\min_{f_{out}, g} \| \mathbf{X}_{in} - \mathbf{Y}_{in} \|^{2} = \min_{f_{out}, g} \| \mathbf{X}_{in} - g(f_{out}(\mathbf{X}_{out})) \|^{2} \quad (2)$$

where $\mathbf{X}_{out}$ is out-of-domain inputs from $D_{out}$. Thus, the objective function of AEDA is formulated as follows:

$$\min_{f_{in}, f_{out}, g} \| \mathbf{X}_{in} - g(f_{in}(\mathbf{X}_{in})) \|^{2} + \| \mathbf{X}_{in} - g(f_{out}(\mathbf{X}_{out})) \|^{2} \quad (3)$$

Neither AE or DAE by itself can guarantee that the hidden layer represents the signals common to both domains. However, the shared decoder $g$ forces the encoders $f_{in}$ and $f_{out}$ to represent the input signal as common to both domains. This constraint is the most significant part of AEDA because we actually do not have the in-domain input $\mathbf{X}_{in}$ that matches the out-of-domain input $\mathbf{X}_{out}$ to estimate DAE part as in Eq. (2). In the visual object recognition area, Jhuo [26] has already demonstrated that out-of-domain samples can be transformed to the in-domain by linear reconstruction of in-domain samples. Thus, the in-domain i-vector $\tilde{\mathbf{X}}_{in}$ that matches $\mathbf{X}_{out}$ is created using sparse reconstruction to obtain a similar distribution to the in-domain i-vector set $D_{in}$. This allows AEDA to be estimated using a sparsely reconstructed $\tilde{\mathbf{X}}_{in}$. Although $\tilde{\mathbf{X}}_{in}$ is not real in-domain data, the encoder $f_{out}$ ensures that the out-of-domain data maps to the common domain by the DAE part and $g$ ensures that the common domain maps to in-domain by the AE part. Without loss of generality, the objective function of AEDA can be formulated as follows:

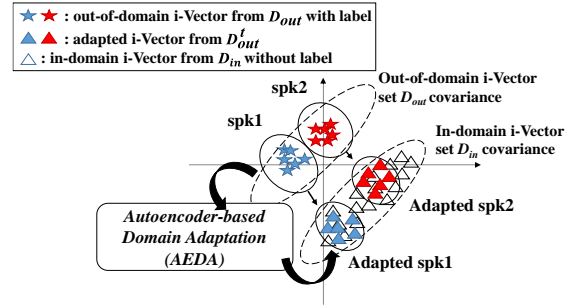

Figure 2: *AEDA approach to adapting out-of-domain dataset to in-domain dataset presented in 2-dimensional space: stars represent out-of-domain dataset and triangles represent in-domain dataset. Red and blue represent speaker labels and white represents no label.*

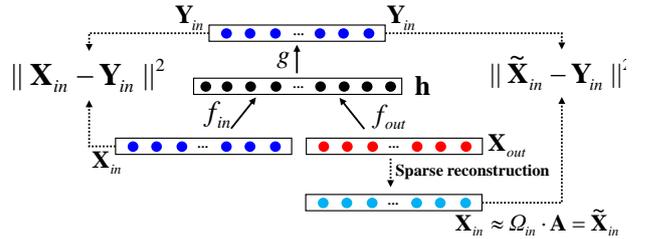

Figure 3: *AEDA architecture, which combines AE (left part) with DAE (right part). Blue and red circles represent in-domain and out-of-domain signals respectively. Black circles represent the hidden representation common to the both domain. Light blue colors represent sparsely reconstructed out-of-domain signals using in-domain dictionary.*

$$\min_{f_{in}, f_{out}, g} \| \mathbf{X}_{in} - g(f_{in}(\mathbf{X}_{in})) \|^{2} + \| \tilde{\mathbf{X}}_{in} - g(f_{out}(\mathbf{X}_{out})) \|^{2}. \quad (4)$$

A sparse representation concept is used to estimate in-domain i-vector $\tilde{\mathbf{X}}_{in}$. In essence, a signal is represented by a linear combination of a small number of basis functions, i.e. a dictionary [26]. Thus, the sparse reconstruction constraint ensures that the reconstructed output follows a similar distribution to the dictionary. $\tilde{\mathbf{X}}_{in}$ can be sparsely reconstructed from the in-domain dictionary matrix $\Omega_{in}$.

$$\Omega_{in} = [\omega_{1}, \omega_{2}, ... \omega_{K}] \in \mathbf{R}^{m \times K} \quad (5)$$

where $\omega_{k}$ is an i-vector from the in-domain dataset $D_{in}$, $K$ is the total number of i-vectors used as a dictionary and $m$ is the dimension of the i-vector. The objective function for reconstruction is

$$\min_{\alpha_{j}} \| \Omega_{in}\alpha_{j} - \mathbf{y}_{j}^{in} \|^{2} + \gamma | \alpha_{j} |^{2} \quad (6)$$

where $\alpha_{j}$ is a coefficient with $K$ elements vector for reconstruction of the i-vector and $\mathbf{y}_{j}^{in} = g(f_{out}(\mathbf{x}_{j}^{out}))$. Here $\mathbf{x}_{j}^{out}$ is $j$-th i-vector from the out-of-domain dataset $D_{out}$ which has $J$ number of utterances in total. $\gamma$ controls for the sparsity of $\alpha_{j}$. From Eq. (6) the in-domain i-vector set can be obtained as

$$\tilde{\mathbf{X}}_{in} = \Omega_{in}\mathbf{A} \quad (7)$$

where $\mathbf{A}$ is the series of $\alpha_{j}$ for all $j$. To optimize AEDA, Eq. (2) should first be initialized by dataset $D_{in}$ and then Eq. (6) and Eq. (4) should be optimized in an iterative manner. Eq. (4) can be optimized by gradient descent as is typical for AE or DAE [19]. Eq. (6) can be optimized by least angle regression [27].

## 4. Experiments and Results

### 4.1. Experimental conditions

For the experiments, DAC 13 i-vector set which consists of SRE and SWB as Table I is used for speaker recognition system as Fig.1. Whitening and length normalization are done by SRE-1phn. The number of eigenvoices of PLDA is set to 400.

Under the proposed AEDA approach, the 1500 i-vectors are selected ($K$=1500) randomly for the in-domain dictionary $\Omega_{in}$. The sparsity $\gamma$ is set to 0.01 and 1000 nodes are used for the hidden layer. Finding the optimal parameters will be explored in a separate study. After constructing the AEDA, the SWB (out-of-domain) dataset is adapted to the SRE-1phn (in-domain) dataset (using $f_{out}$ and $g$) and it is referred to as AEDA-SWB dataset in this paper. Performance is evaluated with the SRE10 test same as in Sec. 2, and the used metrics are EER and the minimum Detection Cost Functions as defined in SRE 2008 (DCF08) and 2010 (DCF10).

### 4.2. Performance comparison to state-of-the-art techniques

The systems' performances according to the three indices are compared in Table III. Systems 3 and 4 are taken from Table II and used as the domain matched and mismatched baselines. State-of-the-art techniques from other studies are examined for comparison. For Garcia-Romero's Interpolated approach [13] referred to as system 5 in Table III, the true speaker label is used for ideal case rather than clustering with AHC algorithm. Then, WC and AC from SWB and SRE-1phn are interpolated, as indicated in Table III by "SWB + SRE-1phn". To obtain the best result, the interpolation parameters for WC and AC are set to 0.6 and 0.3 as in [14]. IDV and DICN performances are presented under systems 6 and 7, respectively. Linear Discriminant Analysis (LDA) is needed for both approaches and i-vectors are projected to 400 dimensional subspace. IDV and DICN approaches are applied on the out-of-domain i-vector set and they are indicated as IDV-SWB and DICN-SWB respectively. For comparison with prior Autoencoder-based method, we developed DAE using out-of-domain dataset which has speaker label with 1300 nodes of single hidden layer as Kudashev's approach [23] and examined as shown in system 8.

Table 3: *SRE10 evaluation result with DAC 13 Dataset when Unlabeled In-Domain Dataset is Available*

| # | Adaptation & Compensation | **WC,AC** | EER | DCF10 | DCF08 |
|---|---|---|---|---|---|
| 3 | - | SRE-1phn | 9.34 | 0.721 | 0.520 |
| 4 | - | SWB | 5.66 | 0.633 | 0.426 |
| 5 | Interpolated [13] | SWB + SRE-1phn | 6.55 | 0.652 | 0.454 |
| 6 | IDV [15] | IDV-SWB | 6.15 | 0.676 | 0.476 |
| 7 | DICN [16] | DICN-SWB | 4.99 | 0.623 | 0.416 |
| 8 | DAE [23] | DAE-SWB | 4.81 | 0.610 | 0.398 |
| 9 | AEDA | AEDA-SWB | **4.50** | **0.589** | **0.362** |

Interpolated approach and the IDV approach show performance degradation under insufficient channel information, and even the domain mismatched system 4 shows better performance than these two approaches. While the performances of IDV and Interpolated approaches lie between those of the baseline systems 3 and 4, DICN and DAE show slight improvement over baselines.

Using the proposed AEDA, in-domain adapted dataset AEDA-SWB can be created from a SWB dataset. This AEDA approach (systems 9) shows better performance than others for all metrics, especially EER, which is improved by 7% over the DAE approach and 20% over the system 4 baseline.

### 4.3. Analysis

Fig. 4 shows the impact of AEDA, which gives better performance than others. The 600-dim LDA is used to represent i-vectors for better discrimination between speakers. It is derived from each SWB, DAE-SWB and AEDA-SWB i-vectors respectively and first 2-dim i-vector are shown in Fig. 4. While both approaches increase cross-speaker variability, the AEDA maintains within-speaker variability at similar level. DAE, on the other hand, fails to maintain within-speaker variability and the speakers overlap each other. This analysis suggests that AEDA could adapt i-vectors from out-of-domain to in-domain successfully.

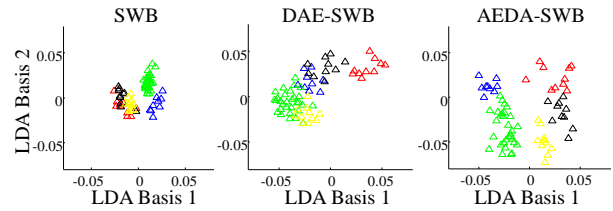

Figure 4: *i-vectors of first five speakers in the SWB dataset using 3 different techniques on first 2 dimension of LDA subspace. Different colors represent different speakers.*

## 5. Conclusions

The recent studies on unsupervised domain adaptation showed performance degradation under insufficient channel information. In this paper, we proposed an AEDA technique which leverages out-of-domain information. In our study, only a small set of unlabeled in-domain i-vectors is used as a dictionary for sparse reconstruction. The proposed AEDA is trained using both out-of-domain and in-domain i-vector sets including a sparsely reconstructed in-domain i-vector set. From the experimental results of the SRE10 test, the system based on our proposed AEDA achieved better performance than other approaches. The experimental result demonstrated that, despite utilizing the dataset with insufficient channel information, the proposed AEDA approach achieves unsupervised domain adaptation effectively and allows the small set of in-domain dataset to be useful for in-domain speaker variability estimation.

## 6. Acknowledgement

This work was supported by the National Research Foundation of Korea (NRF) grant funded by the Korea government (MSIP) (No. 2017R1A2B4012720). This subject is supported by Korea Ministry of Environment (MOE) as "Public Technology Program based on Environmental Policy".